\documentclass[article]{jfm}

\usepackage{graphicx}
\usepackage{newtxtext}
\usepackage{newtxmath}
\usepackage{natbib}
\usepackage{hyperref}
\usepackage{soul}
\hypersetup{
    colorlinks = true,   
    urlcolor   = blue,   
    citecolor  = black,
}

\newcommand{\RomanNumeralCaps}[1]
\linenumbers

\usepackage{amssymb,amsmath}
\usepackage{array}
\usepackage{latexsym}
\usepackage{bezier}
\usepackage{psfrag}
\usepackage{bm}
\usepackage{float}
\usepackage{multirow}
\usepackage{mathrsfs}
\usepackage{color}
\usepackage{marginnote}
\usepackage[pdf]{pstricks}
\usepackage{newtxtext}
\usepackage{newtxmath}
\usepackage[normalem]{ulem}

\newcommand{\Natural}{\mathbb{N}}

\newcommand{\bvec}[1]{\mbox{\bf #1}}

\newcommand{\bsy}[1]{{\boldsymbol{#1}}}

\graphicspath{{./Figs/}}

\shortauthor{}

\title[Chaos in Taylor-Couette flow]{{Mathematically established chaos and forecast of statistics with recurrent patterns in Taylor-Couette flow}}

\author[B.\,Wang,  R.\,Ayats, K.\,Deguchi, A.\,Meseguer \and\, F.\,Mellibovsky]{
  B.\,Wang\aff{1},
  R.\,Ayats\aff{1},
  K.\,Deguchi\aff{2} \corresp{\email{kengo.deguchi@monash.edu}},
  A.\,Meseguer\aff{3} \and\,
  F.\,Mellibovsky\aff{3} \corresp{\email{fernando.mellibovsky@upc.edu}}
}

\affiliation{
  \aff{1} Institute of Science and Technology Austria (ISTA), 3400 Klosterneuburg, Austria
  \aff{2} School of Mathematics, Monash University, VIC 3800, Australia 
  \aff{3} Departament de F{\'\i}sica, Universitat Polit\`ecnica de Catalunya, 08034, Barcelona, Spain}

\begin{document}

\maketitle

\begin{abstract}
The transition to chaos in the subcritical regime of counter-rotating Taylor-Couette flow is investigated using a minimal periodic domain capable of sustaining coherent structures. Following a Feigenbaum cascade, the dynamics are found to be remarkably well approximated by a simple discrete map that admits rigorous proof of its chaotic nature. 
The chaotic set that arises for the map features densely distributed periodic points that are in one-to-one correspondence with unstable periodic orbits (UPOs) of the Navier-Stokes system. 
This supports the increasingly accepted view that UPOs may serve as the backbone of turbulence and, indeed, we demonstrate that it is possible to reconstruct every statistical property of chaotic fluid flow from UPOs.

\end{abstract}

\section{Introduction}

About half a century ago, deterministic chaos theory revolutionised our understanding of why fluid flows are so challenging to predict, with groundbreaking contributions like \citet*{Lo63}. 
Since then, the application of dynamical systems theory to fluid phenomena has developed into a rich and fruitful field \citep*{Ec81,ArFaHa93}. However, a significant divide persists between chaos theory and statistics-based turbulence theory. To help bridge this gap, this paper examines a specific flow pattern in the Taylor-Couette system -- a classic fluid setup extensively studied in theory, experiments, and numerical simulations. 

One of the major challenges of turbulence resides in understanding the mechanisms that produce coherent vortical structures, which low-dimensional models struggle to fully represent. 
With advancements in computational power, dynamical systems theory analyses based on numerically solving the Navier-Stokes equations have gained momentum. In the context of shear flows, research along these lines has {enabled significant progress in the understanding of}
the {physical} mechanisms that govern subcritical transition to turbulence \citep*{Ke05,EcScHo07}. {Invariant solutions, typically in the form of travelling waves or unstable periodic orbits (UPOs), enact a fundamental part}
\citep[see, for example,][]{Na90,ClBu97,KaKi01,ItTo01,Wa03,FaEc03,WeKe04,PrDuKe09,MeMe09,GiHaCv09,DeMeMe14}. 
 
Subsequently, the motivation for finding UPOs shifted towards understanding fully developed turbulence \citep*{KaUhlvanVeen12,BaLo20,GraFlo21,CPTKGS22,YHB21,PNBK24}, reviving Hopf's original idea of using recurrent patterns {\citep{Hopf1948}}. 
UPOs potentially offer a promising path towards comprehending turbulence, as they provide a deterministic representation of coherent structure dynamics. 
However, the challenge of relating the statistical properties of turbulence to UPOs remains an open problem. While recent studies have shown that turbulent trajectories can be approximated by combinations of UPOs \citep{YHB21,PNBK24}, such approximations rely on {\it a posteriori} tuning of weighing parameters.
\textit{Cycle expansion theory} \citep*{AAC90,CCP97} is the only known systematic method for deducing statistical quantities {of a deterministic system exhibiting stochastic dynamics} from {the deterministic properties of its} UPOs, but it has yet to be applied successfully to fluid systems.



Hopf's idea was not immediately accepted because relating deterministic recurrent patterns with the inherently probabilistic nature of turbulence statistics appears rather counterintuitive. However, in simple model dynamical systems, this connection is now well-established through chaos theory, which has advanced considerably since the seminal work of \cite{LiYo75},
who introduced the first mathematical definition of chaos. Among the various definitions of chaos currently in use, Devaney's is the most widely spread \citep{Devaney22} despite being harder to prove than Li-Yorke's \citep{AuKi01}. A key insight from Devaney's definition is that unstable periodic points are densely distributed within the chaotic set. Accordingly, embracing Hopf's proposition is mathematically analogous to approximating real numbers (turbulence) with rational numbers (UPOs). This paper exposes the phenomenon in the most direct way possible for a fluid system.


%

The paper is structured as follows. \S 2 presents the formulation of the Taylor-Couette problem. \S 3 demonstrates that a discrete map, which excellently approximates the flow dynamics, exhibits chaos in the Li-Yorke sense. \S 4 establishes that the map is, in fact, chaotic in the sense of Devaney. In this section, we reconstruct the statistical properties of fluid chaos from those of UPOs, independently of Direct Numerical Simulation (DNS) data. Finally, \S 5 concludes the paper.

\section{Minimal box computation for the Taylor-Couette system
}

\begin{figure}
    \begin{center}
    \begin{tabular}{cc}
      \multirow{2}{*}[2.0cm]{\includegraphics[width=.25\columnwidth]{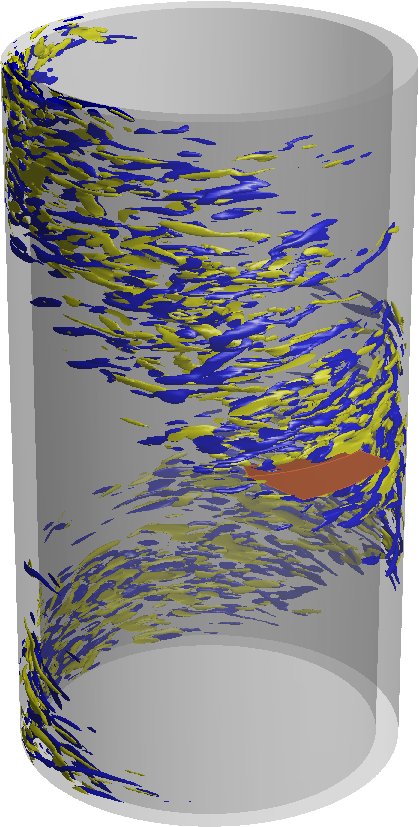}}\put(-104,55){(a)} &
      \includegraphics[width=.35\columnwidth]{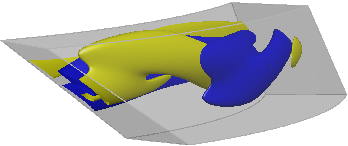}\put(-140,55){(b)} \\[0em] 
      & \includegraphics[width=.35\columnwidth]{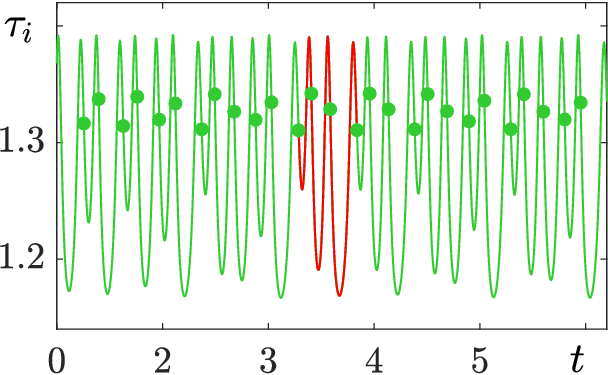}\put(-140,95){(c)}
   \end{tabular} 
   \vspace{5mm}
  \end{center}  
    \caption{Taylor-Couette problem. (a) The spiral turbulence regime visualised by isosurfaces of azimuthal vorticity. The small parallelogram-annular domain in orange is the minimal flow unit used throughout this paper, adopted from \citet{WaAyDe22}. (b) Snapshot of the P$_3$ solution in the minimal flow unit at $R_i=395.7816$. (c) The DNS time signal of inner torque $\tau_i$. Circles denote crossings of the Poincar\'e section $\Sigma$. The red portion indicates a close visit to $\mathrm{P}_3$.}
\label{fig:ProblemSetup1}
\end{figure}

\begin{figure}
    \begin{center}
    \begin{tabular}{cc} 
      \includegraphics[height=0.4\linewidth]{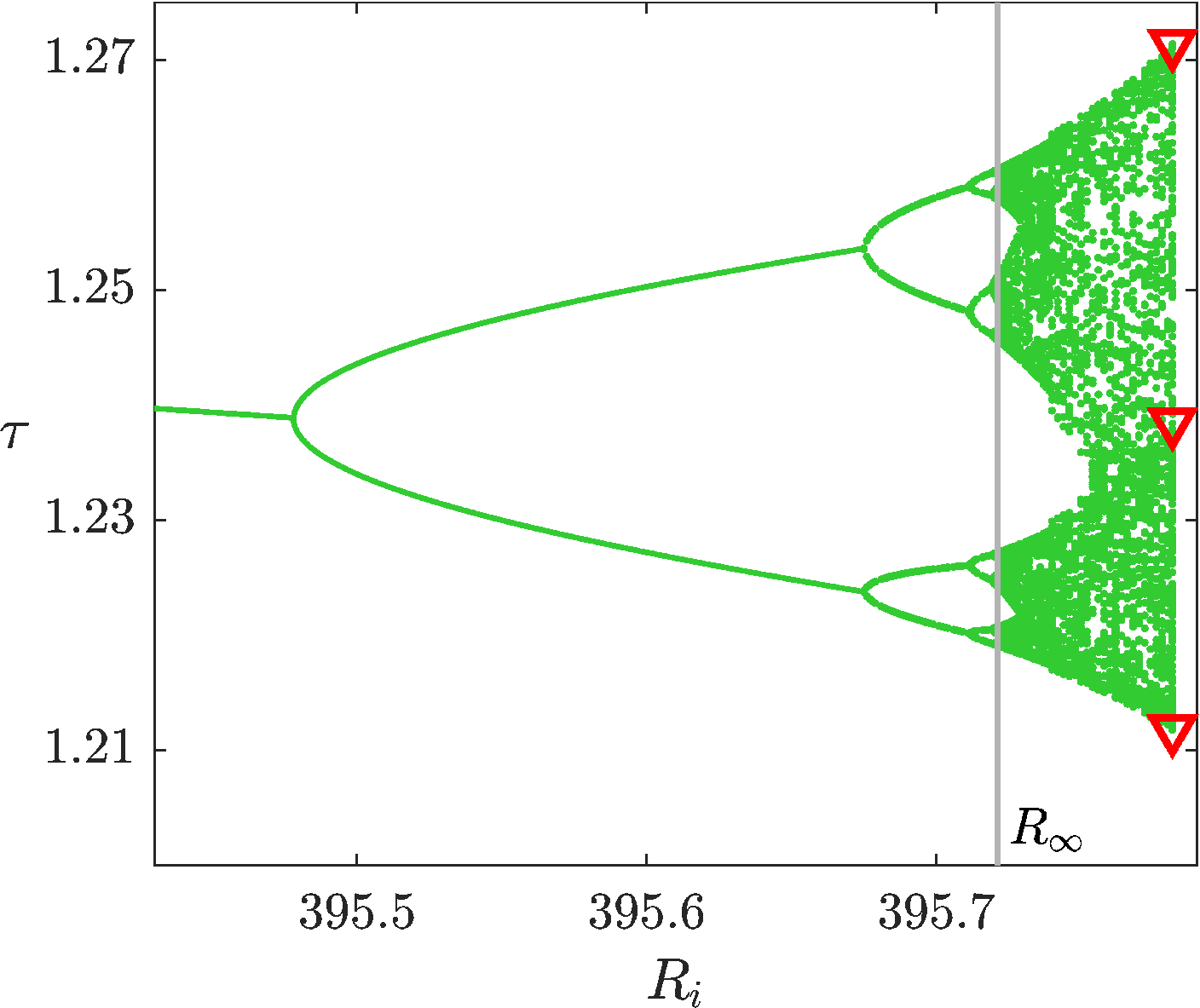} \put(-180,150){(a)} &
      \includegraphics[height=0.41\linewidth]{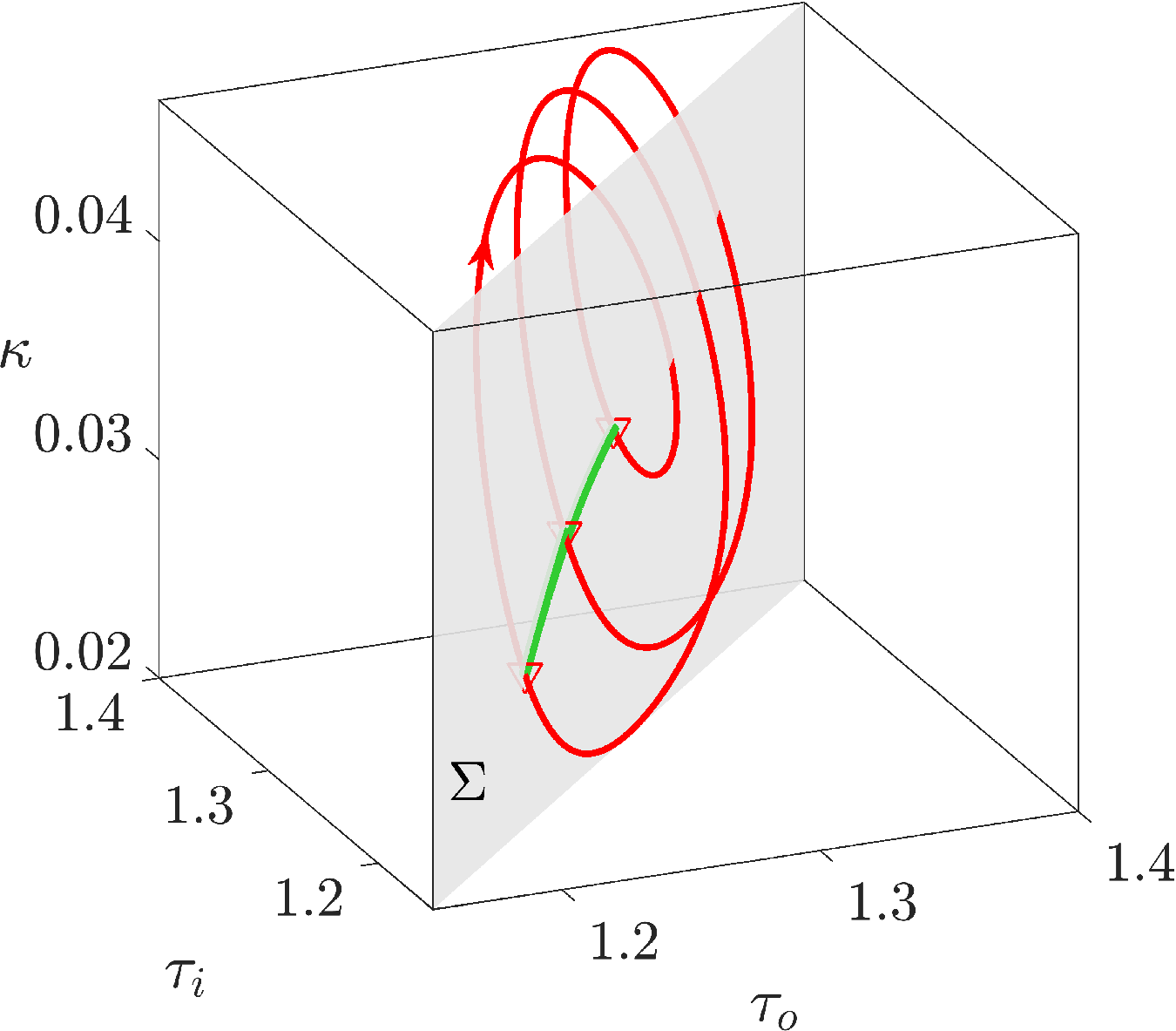} \put(-170,150){(b)}
    \end{tabular}   
  \end{center}  
  \caption{Attractor (from DNS data) and the period-three orbit (from PNK).
  (a) The bifurcation diagram generating the chaotic set. The green dots show torque $\tau$ (on the Poincar\'e section $\Sigma$) as a function of the inner cylinder Reynolds number $R_i$ for statistically steady states in DNS. The red triangles indicate the period-three orbit (P$_3$) at $R_i=395.7816$, at some distance beyond the cascade's accumulation point $R_\infty$.
  (b) Phase map projection on $(\tau_i,\tau_o,\kappa)$ of the P$_3$ orbit (red line) and a representation of the chaotic attractor on the Poincar\'e section $\Sigma$ (green dots) at $R_i=395.7816$. 
  }
\label{fig:ProblemSetup2}
\end{figure}

We target in our investigation the counter-rotating regime of Taylor-Couette flow, known for the subcritical onset of seemingly chaotic dynamics. The emergence of alternating turbulent and laminar helical bands \citep{Co65} has puzzled physicists since its discovery \citep{Fe64}. These peculiar flow structures (figure~\ref{fig:ProblemSetup1}a), which exhibit a clear-cut mean structure \citep{WaMeAy23}, act as harbingers of full-fledged turbulence \citep{AnLiSw86}.
Constraining the turbulent motions within the stripes to small computational domains of parallelogram-annular shape (the small orange box in figure~\ref{fig:ProblemSetup1}a and the domain of figure~\ref{fig:ProblemSetup1}b), which can be regarded as a minimal flow unit \citep{JiMo91,HaKiWa95} for the small scale structures {--about the smallest that can sustain turbulence at sufficiently large rotation rates--} has been recently found to significantly tame the dynamics, thereby  unveiling simple solutions \citep{WaAyDe22}. In this domain,  and for the right set of dynamical parameters, the route to chaos begins with a Feigenbaum cascade \citep{FeigenTC} as shown in figure~\ref{fig:ProblemSetup2}a.
We shall shortly see that, some distance past the accumulation point ($R_\infty$), an unstable period-3 orbit (P$_3$, red triangles) can be identified.

Our computational setup deals with an incompressible fluid of kinematic viscosity $\nu$ confined between two coaxial cylinders of inner and outer radii $r_i$ and $r_o$, respectively. The inner and outer cylinders are independently rotating with respect to their common axis at angular speeds $\Omega_i$ and $\Omega_o$. 
The fluid motion is governed by the Navier-Stokes equations, 
\begin{eqnarray}
\partial_{t}\bvec{v} + (\bvec{v}\cdot\nabla)\bvec{v} = -\nabla (p-zf(t)) + \nabla^{2}\bvec{v},
\end{eqnarray}
expressed in nondimensional form, using the radial gap $d=r_o-r_i$ and $d^2/\nu$ as units of length and time, respectively. This scaling leads to the inner, $R_i=dr_i\Omega_i/\nu$, and outer, $R_o=dr_o\Omega_o/\nu$, Reynolds numbers associated with the rotation of the inner and outer cylinders, respectively.  The pressure field consists of two components: an axial forcing term ($f$) that adjusts instantaneously to keep zero-axial net massflux, and the rest ($p$) that constrains the velocity (${\bf v}$) to comply with the incompressibility condition $\nabla\cdot\bvec{v} = 0$. For convenience, we employ cylindrical coordinates $(r,\theta,z)$. The boundary conditions at the inner and outer cylinder walls $r=r_i$ and $r=r_o$ are $\bvec{v}=R_i\,\hat{\bsy\theta}$ and $\bvec{v}=R_o\,\hat{\bsy\theta}$. Periodicity is enforced to the rest of boundaries of the parallelogram domain (see figure~\ref{fig:ProblemSetup1}b, illustrated with a snapshot of P$_3$). In our particular setup, the radius ratio is $\eta=r_i/r_o=0.883$, and the outer cylinder Reynolds number is fixed at $R_o=-1200$. The transition scenario for these parameter values and increasing inner cylinder Reynolds number $R_i$ has been thoroughly studied \citep{MeMeAv09a,DeMeMe14,WaAyDe22}. Direct numerical simulation (DNS) is performed using a fourth-order linearly implicit (IMEX) scheme, and UPOs are computed with {the} Poincar\'e-Newton-Krylov (PNK) {method}
{detailed in \citet{WaAyDe22} and using the same numerical setup in terms of domain shape, size and resolution.}

We characterise flow states by the inner $\tau_i$ and outer $\tau_o$ cylinder torques. We also use the spatially averaged kinetic energy $\kappa$ of velocity deviation from the laminar circular Couette flow whenever a third observable is necessary. All quantities are normalised by their laminar value, such that $\tau_i=\tau_o=1$ and $\kappa=0$ for circular Couette flow. We use the torques to define the Poincar\'e section as
\begin{equation}\label{eq:PoincSec}
  \Sigma=\left\{\tilde{{\bf v}}\in \mathbb{X}\left |\;\tau_i(\tilde{{\bf v}})=\tau_o({\tilde{\bf v}}),\;\dfrac{d\tau_i}{dt}>\dfrac{d\tau_o}{dt}\right. \right\},
\end{equation}
where $\tilde{\mathbf{v}}$ is the velocity field $\mathbf{v}$, duly shifted with the method of slices \citep{WiCvAv13,WaMeAy23} to remove the degeneracy induced by the spatial drift of the solutions{, both in the axial and azimuthal directions}, and $\mathbb{X}$ is the corresponding phase space. The bifurcation diagram shown in figure~\ref{fig:ProblemSetup2}a records the torque, on $\Sigma$, of statistically steady states. Beyond the accumulation point of the cascade, $R_{\infty}$, the seemingly chaotic attractor forms banded structures in the bifurcation diagram that progressively widen and successively merge in pairs.

Hereafter we will focus on $R_i=395.7816$. The energy and torque DNS signals occasionally produce nearly periodic timestamps at this value of the parameter (figure~\ref{fig:ProblemSetup1}c). States taken from selected pseudo-periodic lapses converge onto UPOs when fed to the PNK scheme. A phase map projection of one such UPO is shown in figure~\ref{fig:ProblemSetup2}b.
The trajectory pierces three times the Poincar\'e section every period, hence our naming it P$_3$ (see figure~\ref{fig:ProblemSetup1}b for a snapshot of P$_3$ on $\Sigma$).

\section{P$_3$ implies chaos}

It is well-known in 
chaos theory that for 1d maps generated by continuous functions ${f\!: I\rightarrow I}$, where ${I\subset\mathbb{R}}$ is an interval, the advent of period-3 orbits holds a special significance. Specifically, as established by \citet{LiYo75}, it implies chaos in the sense that the map enhances a swift mixing of the points in the interval. Moreover, according to Sharkovskii's theorem \citep{Sharkovskii64}, the presence of a period-3 point entails also the existence of infinitely many periodic points. 

The presence of P$_3$ in our system therefore renders the application of the Li-Yorke and Sharkovskii theorems highly enticing. However, the theorems hinge upon an imperative condition: the map must be \textit{one-dimensional} and \textit{continuous}. These theorems might not hold true for higher dimensional systems, even if period-3 orbits were in existence. While instances of period-3 solutions abound in the fluid dynamics literature \citep{MoToKn83,KoMoToWe86,KrEc12}, no study has positively checked, to the best of our knowledge, the conditions for the validity of these theorems.

\begin{figure*}                  
  \begin{center}
    \begin{tabular}{cc} 
      \includegraphics[height=0.4\linewidth]{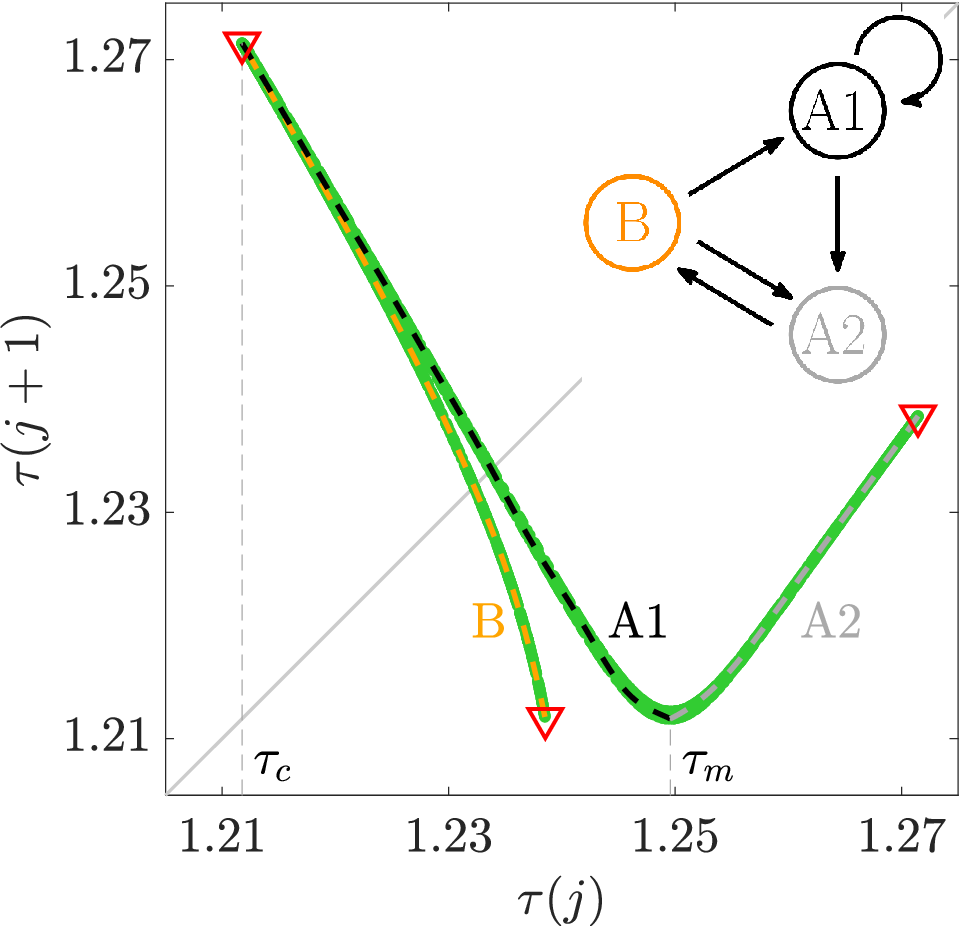} \put(-165,150){(a)} &
      \includegraphics[height=0.41\linewidth]{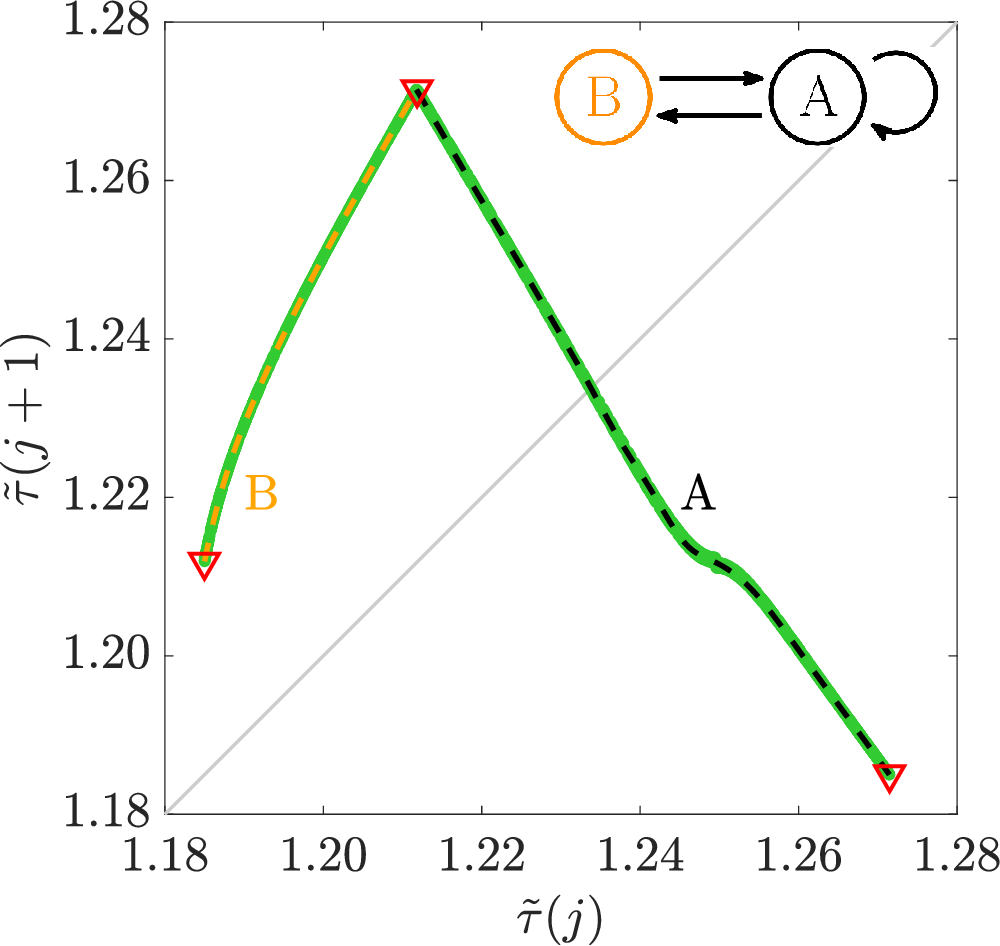} \put(-170,150){(b)}
    \end{tabular}
  \end{center}  
  \caption{Return map analyses based on the Poincar\'e map on $\Sigma$ at $R_i=395.7816$. 
  (a) Return map of the chaotic attractor (green dots) and of P$_3$ (red triangles). The cusp ($\tau_c$) and minimum ($\tau_m$) points split the map in three distinct branches: B (orange), A1 (black) and A2 (gray). (b) The same return map, now unfolded following \eqref{tautilde}. The A1 and A2 branches are now labelled as A (black). The inset diagrams indicate branch selection rules.
  }
  \label{fig:P3ChaoticAttractor}
\end{figure*}
A very long DNS (over 500 viscous time units) was conducted, with the solution $\tilde{{\bf v}}(t_j)$ recorded at the chronologically ordered discrete times $t_j$, $j=0,1,2,\dots$, at which the phase map trajectory crosses the Poincar\'e section $\Sigma$. The approximately 2500 data points we collected are shown as green dots in figure~\ref{fig:ProblemSetup2}b. All points are tightly clustered on a very narrow band, akin to a one-dimensional curve. This holds for any projection of $\mathbb{X}$ one might choose, which indicates a highly restricted dimensionality of the first return (Poincar\'e) map $\Phi: \Sigma \longrightarrow \Sigma$. Strictly speaking, the dimension cannot be exactly one in the case of reversible maps but, as demonstrated by well-known examples like the H\'enon map or the Lorenz system, it can approach one. Indeed, since our system is at a relatively low Reynolds number, it is plausible that the dimension of the inertial manifold \citep{Te89,Di16,HaKaLiAx23} might be particularly low. Figure~\ref{fig:P3ChaoticAttractor}a shows the return map in terms of the torque $\tau(j)\equiv\tau_i(\tilde{{\bf v}}(t_j))=\tau_o(\tilde{{\bf v}}(t_j))$. The multivaluedness of the map hinders direct application of standard 1d dynamical systems theory, but this can be readily addressed by simple identification of the selection rule among branches. The curve in figure~\ref{fig:P3ChaoticAttractor}a is naturally split at the cusp point $\tau=\tau_c$ into two main branches: A (gray scale) and B (orange). Branch A is then further split in two sub-branches, A1 (dark) and A2 (light), for convenience. The graph in the inset summarises the selection rule among branches. A point on branch A1 might be mapped onto itself or sent to branch A2 and thence immediately to branch B. Then branch B instantly repels the trajectory, but whether A1 or A2 is to follow depends on the landing location along the branch.

A single-valued version of the map can then be constructed by applying the change of variable
\begin{equation}\label{tautilde}
  \tilde{\tau}(j) =
  \left\{
  \begin{array}{ll}
    2\tau_c-\tau(j) & \text{if $\tau(j-1)>\tau_m$} \\
    \tau(j) & \mathrm{otherwise}
  \end{array}
  \right.,
\end{equation}
where $\tau_m \approx 1.250$ is the location of the local minimum of the multivalued function, whose immediate surroundings are mapped to the vicinity of $\tau_c \approx 1.212$. This change of variable corresponds effectively to the unfolding of the map, shown in figure~\ref{fig:P3ChaoticAttractor}b, which now admits standard cob-webbing to monitor the dynamics. By interpolating the DNS data points and P$_3$ using splines, it is possible to construct a continuous function $S:I \rightarrow I$, where $I\approx[1.185,1.271]$. 
The spline dynamical system $\tilde{\tau}(j+1)=S(\tilde{\tau}(j))$ is chaotic in the Li-Yorke sense, because it contains the P$_3$ points, which in this case appear at the vertex and endpoints of the continuous function $S$. 

\begin{figure}
  \begin{center}
    \raisebox{0.2em}{\includegraphics[width=0.65\linewidth]{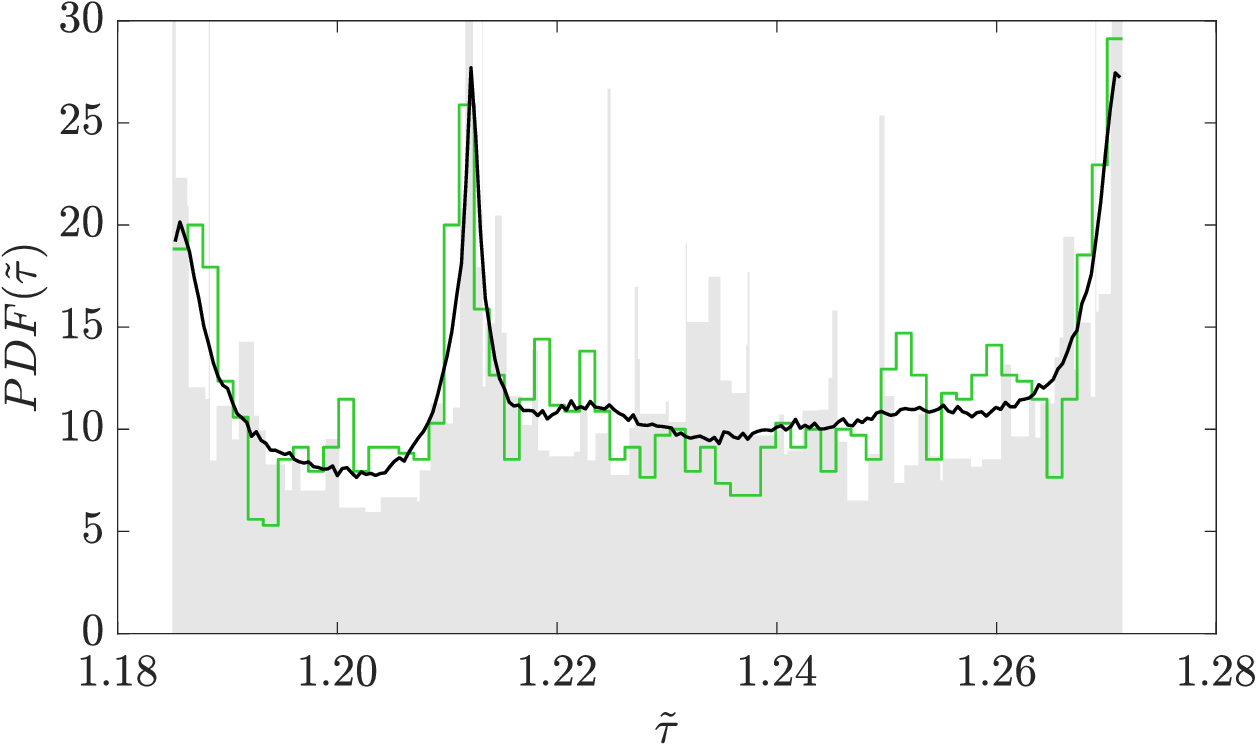}}
  \end{center}  
  \caption{Probability density function (PDF) for the spline dynamical system (black), the normalised histogram for DNS data of the chaotic attractor (green), and the prediction from periodic points (gray). 
  }
  \label{fig:PDF}
\end{figure}

A natural question follows regarding the extent to which iteration of the spline map reproduces DNS results, an issue that needs assessing. To this end, we compare the probability density functions (PDF) in figure~\ref{fig:PDF}.
The PDF corresponding to the spline dynamical system (black curve), with its Lyapunov exponent of about $0.47$, excellently matches DNS results (green {curve}). The emergence of peaks in the PDF around the period-3 points is a common feature of the intermittency route to chaos \citep{PoMa80,Ec81}.

\section{Probability density function computed from unstable periodic solutions}
\begin{figure}

    \begin{center}
    \includegraphics[width=0.65\linewidth]{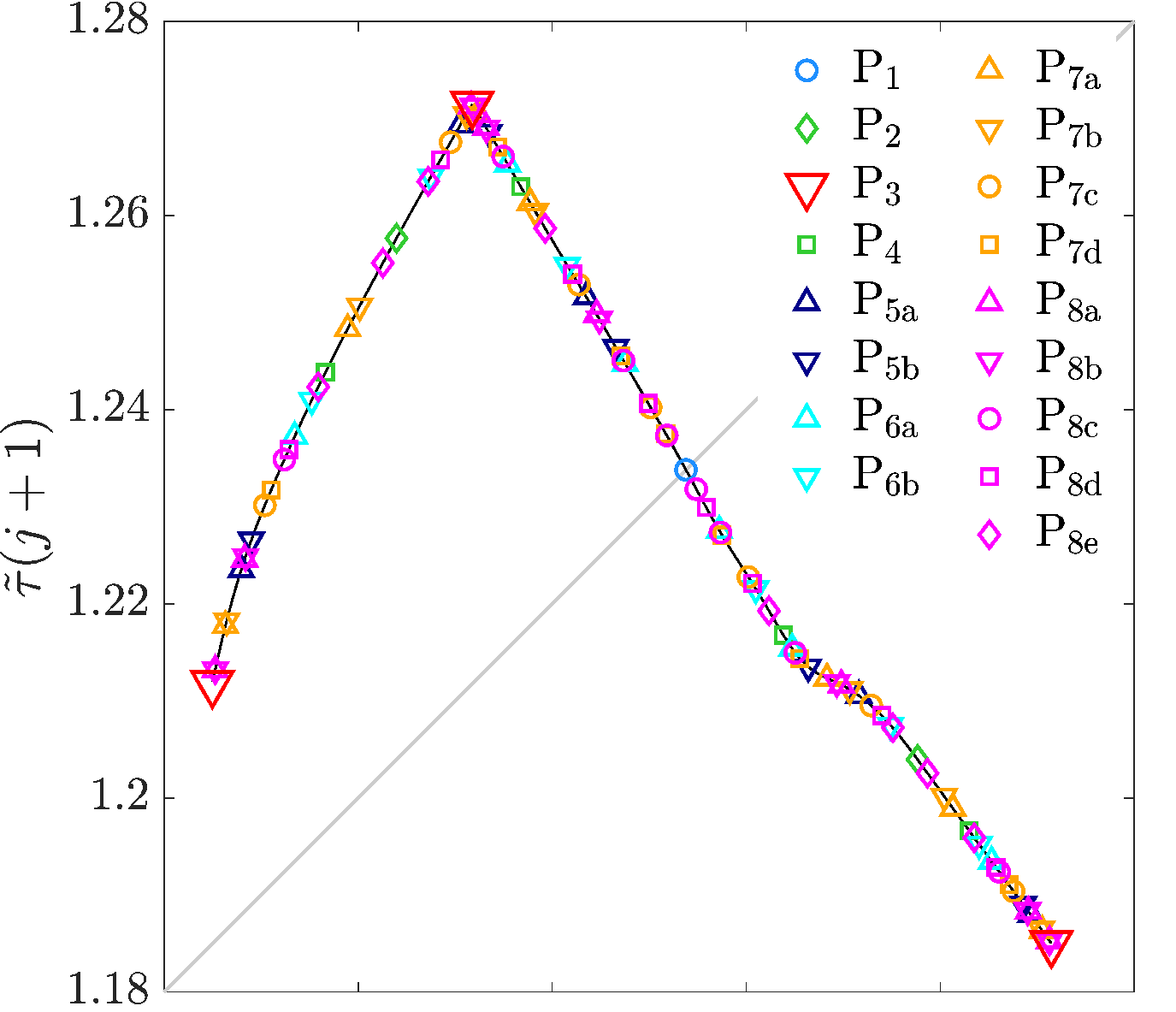}\put(-270,220){(a)} \\
    \includegraphics[width=0.65\linewidth]{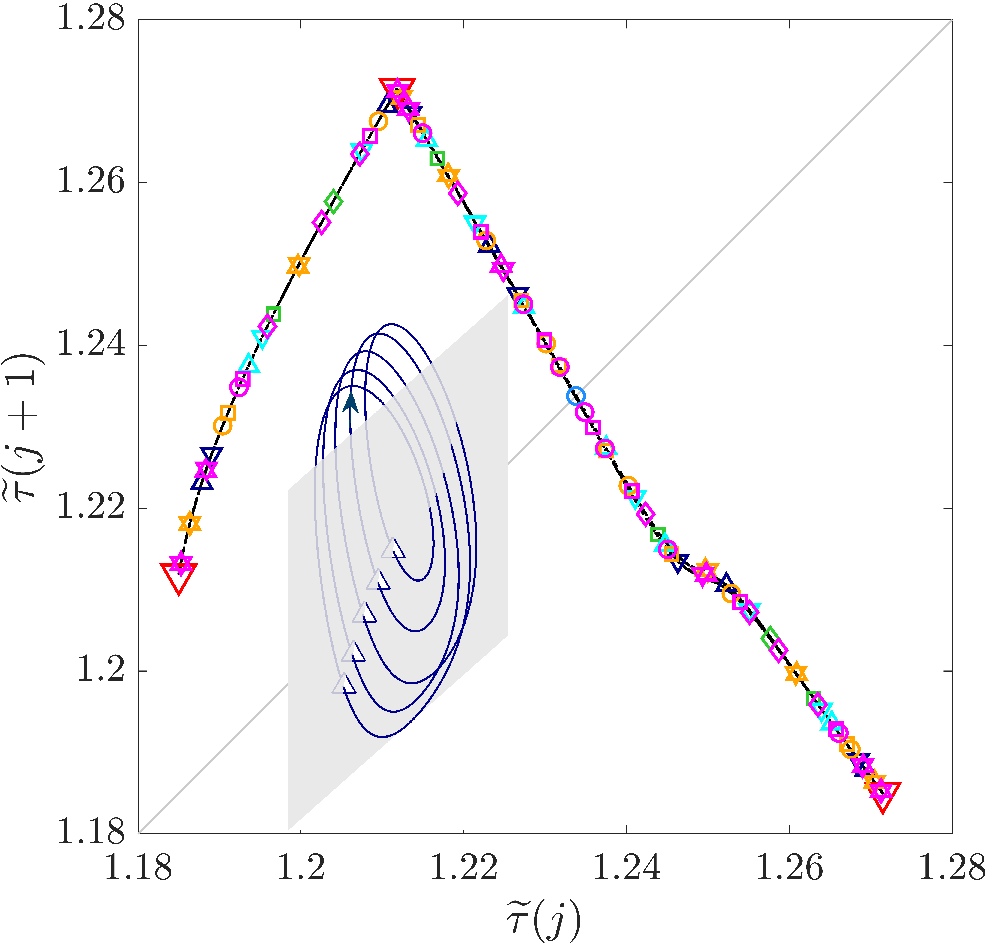}\put(-270,240){(b)}
  \end{center}  
 
\caption{Complete set of periodic points up to period $n=8$ for (a) the spline map approximation and (b) the Navier-Stokes system. The inset shows a phase map projection of P$_{5a}$ analogous to that of P$_3$ in figure ~\ref{fig:P3ChaoticAttractor}a.
}
  \label{fig:SplineNSRetMap}
\end{figure}

The inherent relation between the spline dynamical system and the Navier-Stokes dynamics becomes all the more evident through the comparison of periodic solutions. For the 1d spline map, Sharkovskii's theorem guarantees the existence of periodic points of period $n$ for every $n \in \Natural$. Figure~\ref{fig:SplineNSRetMap}a shows the complete set up to $n=8$.
We have been able to find the Navier-Stokes counterpart to each and every one of the spline periodic points employing PNK (see figure~\ref{fig:SplineNSRetMap}b). The discrepancy in terms of torque between spline map and Navier-Stokes solutions is consistently below $0.25\%$.

The fluid flow scenario we have chosen is accurately described by a map of great mathematical simplicity, whose elegant attributes 
{make it possible to infer the PDF of chaotic dynamics from the underlying periodic points in a clear and straightforward manner.}
First we note that the branch selection of the unfolded map, shown in the inset of figure~\ref{fig:P3ChaoticAttractor}b, is extremely simple, with B and A representing the branches to the left and right of the maximum, respectively. The diagram is nothing but the topological Markov chain over the alphabet $\{A, B\}$, whose topological entropy is known to be the natural logarithm of the golden ratio $\phi=(1+\sqrt{5})/2$ (see \cite{CAMTV16} for the derivation). The spline map being an interval map of strictly positive entropy, one can conclude that it must be chaotic in the sense of Devaney \citep{Li93,Devaney22}.
The periodic points are, therefore, densely distributed within the chaotic set, which supports the recurring assertion among fluid dynamicists that UPOs form the skeleton of turbulence \citep{KaUhlvanVeen12,GraFlo21,AvBaHo23,CPTKGS22}. The 100 periodic points we have computed cover the map effectively, thus suggesting that UPOs indeed encode the dynamics and, consequently, hold the key to predicting the statistical properties inherent to chaos.

\begin{figure}
    \begin{center}
    \begin{tabular}{cc}
      \includegraphics[width=0.5\linewidth]{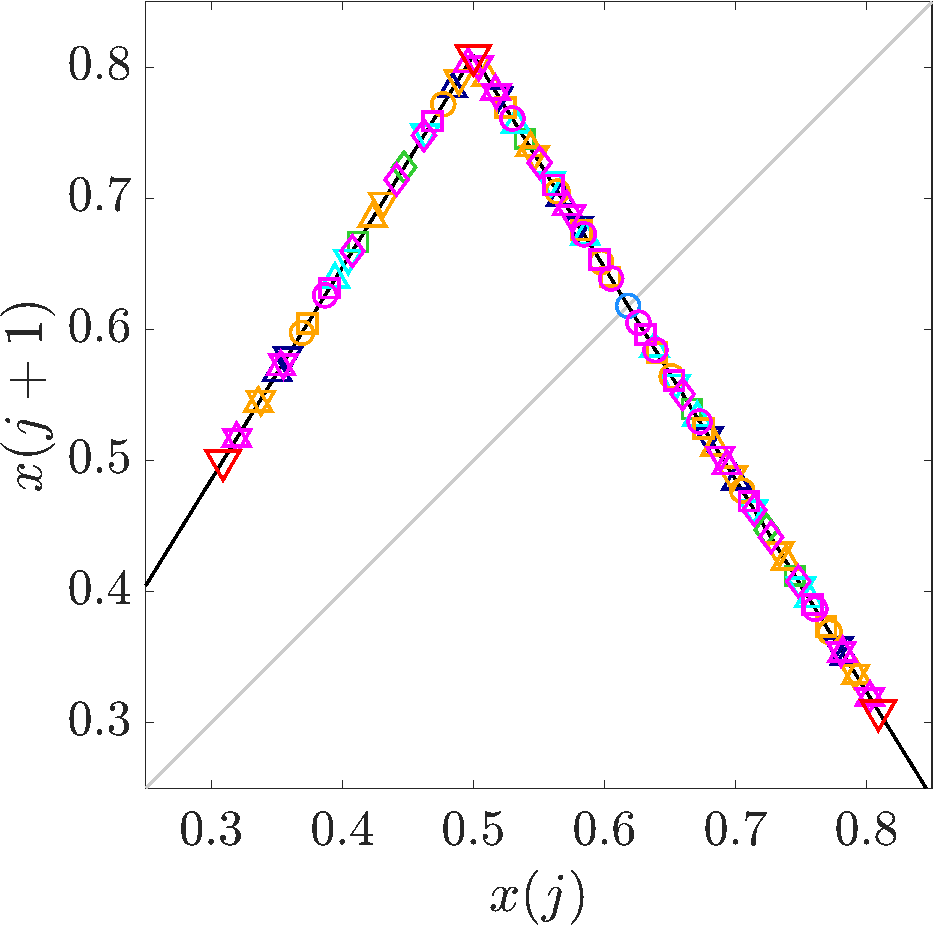} &
      \raisebox{0.3em}{\includegraphics[width=0.265\linewidth]{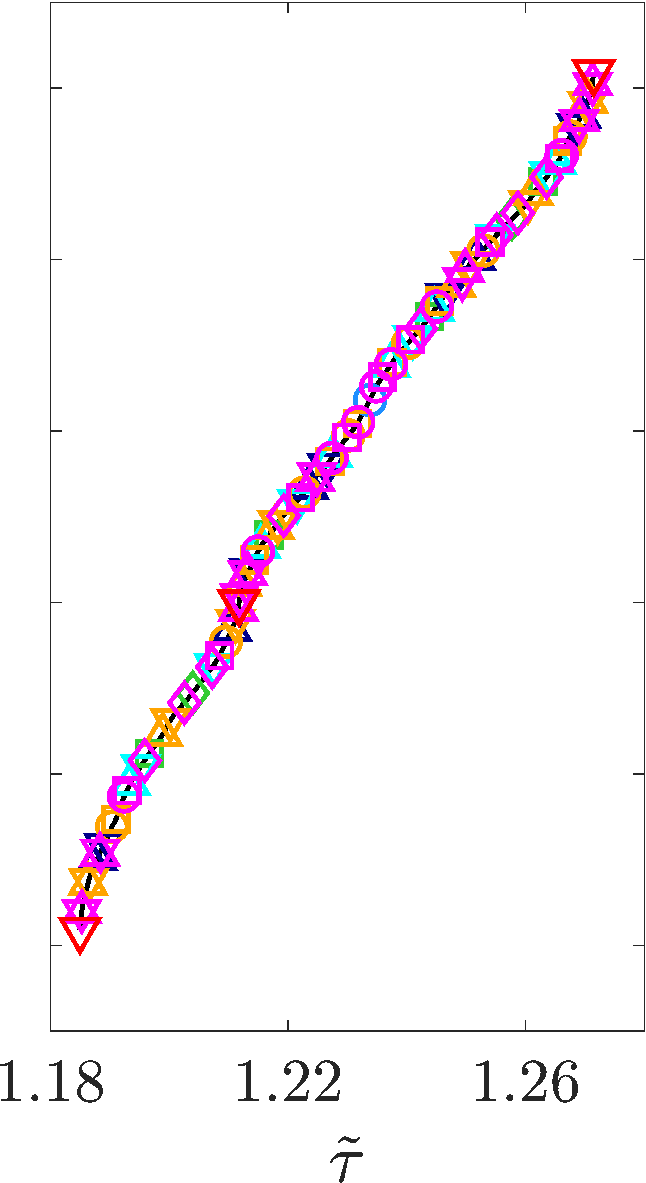}}
      \put(-300,175){(a)}\put(-107,175){(b)}
    \end{tabular}
  \end{center}  
  \caption{Topological conjugacy of the tent and spline maps. (a) The tent map $T(x)$ with all periodic points up to period $n\leq 8$. (b) Correspondence between tent map (ordinate) and spline map (abscissa) periodic points. The piecewise linear curve connecting the points provides an approximation of the conjugacy homeomorphism $h(\tilde{\tau})$. Symbols and colours, representing periodic orbits, follow the legend used in previous figures.}
  \label{fig:TentMap}
\end{figure}

The PDF of the spline map can be obtained from that of the tent map, $T(x)=\phi(1-|2x-1|)/2$,
with the scale parameter tuned to induce the appropriate topological Markov chain{, i.e. having the same topological entropy $\phi$}. Its PDF has a piecewise-constant closed analytical expression \citep{CAMTV16}:
\begin{eqnarray} 
  \rho_{T}(x)=
  \left \{
  \begin{array}{ccr}
    \dfrac{2\phi}{2\phi-1} & \text{if} & \dfrac{\phi-1}{2}\leq x<\dfrac{1}{2},\\[1em]
    \dfrac{2\phi}{3-\phi} & \text{if} & \dfrac{1}{2} \leq x \leq \dfrac{\phi}{2}.
  \end{array}
  \right.
\end{eqnarray}
Remarkably, the tent $T$ and spline $S$ maps share a common list of periodic solutions, appearing in the exact same order along the curve (figure~\ref{fig:TentMap}a).
This implies the existence of a homeomorphism $h$ such that $h(S(\tilde{\tau}))=T(h(\tilde{\tau}))$, i.e., the two maps are topologically conjugate in the sense that they are dynamically equivalent \citep{YYCOP00}. A discrete sampling of $h$ emerges naturally when plotting corresponding periodic points of the two maps on the $\tilde{\tau}$--$x$ plane (figure~\ref{fig:TentMap}b). A straightforward calculation shows that the PDF of the spline and tent maps are related by $\rho_{S}(\tilde{\tau})=\rho_{T}(h(\tilde{\tau}))|h'(\tilde{\tau})|$,
where the prime denotes ordinary differentiation. The prediction, the gray shaded region shown in figure~\ref{fig:PDF}, is obtained from a piecewise linear approximation of $h$ based on the available periodic points. The agreement with the spline and DNS data is fair despite the roughness of the estimation, as attested by the accurate reproduction of the three salient peaks. 

The prediction error in the PDF primarily occurs at high frequencies, thus having a small impact on the lower-order statistical moments. The expectation $E$ and variance $V$ of the torque computed from the PDFs are $(E,V)=(1.2294,6.9341 \times 10^{-4}),(1.2288,7.1341 \times 10^{-4})$ and $(1.2289,6.9144 \times 10^{-4})$, for the spline dynamical system, DNS, and the UPOs prediction, respectively.
Replacing the spline periodic points by the Navier-Stokes UPOs yields nearly the same results, as expected from the excellent agreement exhibited by figure~\ref{fig:SplineNSRetMap}. Furthermore, the one-dimensional assumption of the return map implies that the PDF for the full velocity field $\tilde{\mathbf{v}}$ can be determined in a similar manner. Therefore, all widely used turbulence statistics can, in principle, be computed using UPOs.

\section{Conclusion}
We have analysed, in the Taylor-Couette system, a subcritical parameter regime exhibiting 
dynamics that can be approximated by a simple discrete map. The map has exceptionally neat mathematical properties, which allow to rigorously establish its chaotic nature as well as the existence of infinitely many unstable periodic orbits. Remarkably, the fluid system and the discrete map share a common catalog of unstable periodic solutions with the tent map, a clear indication of topological conjugacy. A sufficient number of these solutions enables the construction of a conjugacy homeomorphism, which can be used to predict the probability density function to be expected from direct numerical simulation. 

Our informed choice of scenario purposely aims at greatly simplifying the theoretical analysis. For more general cases, the topological properties of chaotic fluid systems could be approximated by simple Markov maps \citep{YHB21}. As demonstrated in this paper, UPOs serve precisely as the link between such Markov maps and the Navier-Stokes system. Our theory differs from cycle expansion theory \citep{AAC90,CE91,CCP97,CAMTV16} and offers a more intuitive understanding of how the PDF can be constructed from UPOs. On the other hand, cycle expansion theory has the advantage of permitting the computation of statistically averaged quantities in a simpler manner by applying prescribed formulas.

A complete understanding of turbulence is still a long way off from both the applied and the mathematical stand points. Whether Hopf's idea can be extended to high Reynolds number turbulence remains an open question. 
Furthermore, our results, depending on numerical approximations, do not guarantee that the Navier-Stokes solutions exhibit chaotic behaviour in the sense of Devaney.
Nonetheless, we expect that our demonstration of how chaos theory applies to fluid flows under certain conditions will encourage future research on this important yet difficult problem.


\backsection[Acknowledgements]{
This research is supported by the Australian Research Council Discovery Project DP230102188 and the Ministerio de Ciencia, Innovación y Universidades (Agencia Estatal de Investigación, project nos. PID2020-114043GB-I00 (MCIN/AEI/10.13039/501100011033) and PID2023-150029NB-I00 (MCIN/AEI/10.13039 /501100011033/FEDER, UE). BW's and RA's research has been funded by the European Union’s Horizon 2020 research and innovation programme (Marie Sk\l{}odowska-Curie Grant Agreement No. 101034413). RA has also been funded by the Austrian Science Fund (FWF) 10.55776/ESP1481224.}

\backsection[Author ORCIDs]{B. Wang, https://orcid.org/0000-0002-6229-0336; R. Ayats, https://orcid.org/0000-0001-6572-0621; K. Deguchi, https://orcid.org/0000-0002-3709-3242; A. Meseguer, https://orcid.org/0000-0002-2022-2001; F. Mellibovsky, https://orcid.org/0000-0003-0497-9052}

\backsection[Declaration of Interests]{
The authors report no conflict of interest.
}

\bibliography{local}
\bibliographystyle{jfm}

\end{document}